\documentclass[twocolumn,showpacs,preprintnumbers,amsmath,amssymb]{revtex4}
\begin{document}

\preprint{ }

\title{ 6-dimensional Kaluza-Klein Theory for Basic Quantum Particles and Electron-Photon Interaction}  
\author{\normalsize Xiaodong Chen}
\altaffiliation{2073B Vestavia Park Ct, Birmingham, AL 35216, U.S.A.}
\email{xiaodong.chen@gmail.com} 

\date{\today}

\begin{abstract} 
By extending original Kaluza-Klein theory to 6-dimension, the basic quantum field equations for 0-spin particle,
1-spin particle and 1/2 spin particle with mass $>$0 are directly derived from 6-dimensional Einstein equations. 
It shows that the current quantum field
equations of basic particles become pure geometry properties under 6-dimension time-space. The field equations
of electron and photon can be unified in one 6-dimensional extended Maxwell equation. 
The equations containing interactions between electron and photon will be derived from Einstein equation 
under 6-dimension time-space. It shows that the interactions in QED can be considered as
the effect of local geometry curvature changing instead of exchange virtual photons. 

\end{abstract}

\maketitle

\section{Introduction} \label{INTRO}

Kaluza \cite{Kal21} first developed his method in 1919 in an attempt to unify Electromagnetism 
and General Relativity. In fact, Kaluza's idea showed the possibility to unify not only
gravity and electromagnetism, but also matter and geometry \cite{Overduin}. It gave physicists the hope of extending
Einstein's vision of nature as pure geometry to quantum level. The original Kaluza metric can be 
written as follows: 
\begin{equation}
\left( \hat{g}_{AB} \right) = \left( \begin{array}{cc}
   g_{\alpha\beta} - \phi A_{\alpha} A_{\beta} \; \; & \; \; 
     -\phi A_{\alpha} \\
   - \phi A_{\beta} \; \;                                & \; \; 
     - \phi
   \end{array} \right) \; \; \; 
\label{5dMetric}
\end{equation}
where the $\alpha\beta$-part of $\hat{g}_{AB}$ with $g_{\alpha\beta}$ 
(the four-dimensional metric tensor), the $\alpha 4$-part with $A_{\alpha}$ (the electromagnetic potential),
and the $44$-part with $\phi$  (a scalar field). The four-dimensional metric signature is taken 
to be $(+ \, - \, - \, -)$

The metric above contains a massless scalar field and a massless 1-spin vector field. The method did not
include the mass part of quantum field (although people believes that the metric (\ref{5dMetric}) describe 
a spin-2 graviton, 
a spin-1 photon and a spin-0 boson which is thought to be connected with how particles acquire mass.);
it also can not be used to describe the particle with half integer spin. To completely extend Einstein's vision
of nature (as pure geometry) to quantum level, we need include quantum fields for particles of
 mass $>0$, and the particles of $\frac{1}{2}$ spin into the theory. In other efforts, a possible 
``theory of everything''-- Superstring theory extended Kaluza's idea to ten-dimension,  
 which contains all kinds of fields. But superstring theory is much complicate, it lost the beauty of 
 simplicity in original Kaluza-Klein theory. 
 
In this paper, one will find that by extending Kaluza-Klein theory to 6-dimension time-space, 
the field equations of 0-spin particle with mass $>0$ (Klein-Gordon equation), equations of 1-spin particle with 
mass $>0$ and equations of half integer spin particle (Dirac equation) can be derived directly from 6-dimensional 
Einstein equations under new 6-dimensional metric, especially the equations of electron and photon can be unified 
in one 6-dimensional Maxwell equation when we expand the components of fields of electron and photon 
to 6-dimension. 
It indicates that the current quantum field
equations of basic particles are pure geometry properties of 6-dimension time-space. 
The coupling equations between electron and photon will be obtained from Einstein equation under 6-dimension 
time-space. It shows that the interactions in QED can be considered as
the effect of local geometry curvature changing instead of exchange virtual photons.

\section{Equations of 0-spin free particle} \label{ZeroSpin}

Throughout this paper, the four-dimensional metric signature is taken to be $(+ \, - \, - \, -)$, 
and the indices for 6-dimensional time-space to be 0,1,2,3,4,5. 

The 6-dimensional Einstein equations keep the same format as 4-dimensional:
\begin{equation}
\hat{G}_{AB} = \kappa \hat{T}{AB} \; \; \; ,
\label{5dEFE1}
\end{equation}
where $\hat{T}{AB}$ is 6-dimensional energy momentum tensor, 
$\hat{G}_{AB} \equiv \hat{R}_{AB} - \hat{R} \, \hat{g}_{AB} / 2$
is the Einstein tensor, $\hat{R}_{AB}$ and
$\hat{R} = \hat{g}_{AB} \hat{R}^{AB}$ are 
the 6-dimensional Ricci tensor and scalar respectively,
and $\hat{g}_{AB}$ is the 6-dimensional metric tensor, A, B.. run over 0,1,2,3,4,5 .

The 6-dimensional Ricci tensor and Christoffel symbols are defined
in terms of the metric exactly as in four dimensions:
\begin{eqnarray}
\hat{R}_{AB}        & = & \partial_C \hat{\Gamma}^C_{AB} -
                          \partial_B \hat{\Gamma}^C_{AC} +
                          \hat{\Gamma}^C_{AB} \hat{\Gamma}^D_{CD} -
                          \hat{\Gamma}^C_{AD} \hat{\Gamma}^D_{BC} 
                          \; \; \; , \nonumber \\
\hat{\Gamma}^C_{AB} & = & \frac{1}{2} \hat{g}^{CD} \left( 
                          \partial_A \hat{g}_{DB} +
                          \partial_B \hat{g}_{DA} -
                          \partial_D \hat{g}_{AB} \right) \; \; \; 
\label{6dChristRicci}
\end{eqnarray}
where A, B.. run over 0,1,2,3,4,5. 

For 0-spin particle, we choose the 6-dimensional time-space metric as:
\begin{equation}
\left( \hat{g}_{AB} \right) = \left( \begin{array}{cc}
   g_{\alpha\beta} \; \; \; \; \; \;  \; \; \; \; \; \; \\
  \; \; \; \; \; \; \; \; \phi \; \; \; \; \; \; \; \; \\
   \; \; \; \;  \; \; \; \; \; \; \; \; \; \; -1 \\ 
   \end{array} \right) 
\label{6dMetric_0}
\end{equation}
where metric elements $g_{\alpha\beta}$ is 4-dimensional metric. We concentrate on quantum field equations in
this paper, so ignore gravity field, then 
$g_{\alpha\beta} = {\delta}_{\alpha\beta}$, $g_{55} = -1$, $g_{44} = \phi$,
with conditions: 
\begin{eqnarray}
\partial_{4} \phi = 0  \;, \; \;  \;
\partial_{5} \phi =  i\frac{m_0}{\hbar} \phi \; \; \;  \; 
\label{6DCondition} 
\end{eqnarray}
where $\hbar$ is Planck constant, $m_0$ is rest mass of particle. or equivalently
\begin{equation}
\phi =  \phi_k e^{\frac{im_0}{\hbar} x_5} 
\end{equation}
where $\phi_k$ is $\phi$ in Kaluza metric (\ref{5dMetric}). Start from here,
throughout this paper, capital Latin indices 
A,B,C .. run over 0,1,2,3,5 (A,B $<> 4$),
Greek indices $\alpha, \beta$ ... run over 0, 1, 2, 3, and small Latin indices a, b, ... run over 1, 2, 3.
Also we have $g_{AB} g^{AB} = \delta_{AB} $, that makes $g^{44} = \frac{1}{\phi} $.

Using equations (\ref{6dMetric_0}), (\ref{6dChristRicci}), the $\alpha \beta-$, $\alpha 5- $, 55-, $\alpha 4- $, 
and 44-components of Einstein equations (\ref{5dEFE1}) become:
\begin{equation}
 \frac{1}{\phi}\partial^{\alpha}\partial_{\beta}\phi = \kappa \hat{T}{\alpha\beta} 
\label{5dMSTensor_0}
\end{equation}
\begin{eqnarray}
\frac{-im_0}{\hbar \phi} \partial_{\alpha} \phi = \kappa \hat{T}{5\alpha} = \kappa \hat{T}{\alpha 5} \\
\label{5dMSTensor_1}
-(\frac{m_0}{\hbar})^2 = \kappa \hat{T}{55}     \\
\label{5dMSTensor_2}
 \kappa \hat{T}{4 \beta} = \kappa \hat{T}{\alpha 4} = 0 \\
\label{5dSETensor_3}
\Box \phi + (\frac{m_0}{\hbar})^2\phi = 0  
\label{5dSETensor_4}
\end{eqnarray}
Here $\hat{T}{44} = 0$, i.e. no 5th dimensional energy momentum tensor. Equation (\ref{5dSETensor_4}) 
is Klein-Gordon Equation for free 0-spin particle with mass $>0$. Notice that for free {\it single} particle:
\begin{equation}
\phi = e^{\frac{-i}{\hbar}(p^\alpha x_{\alpha} - m_0 x_5)}
\label{plan_wave}
\end{equation}
Then equation (\ref{5dMSTensor_0}) become: 
\begin{equation}
 -\frac{1}{\hbar^2}p^{\alpha}p_{\beta} = \kappa \hat{T}{\alpha\beta}
\label{5dMSTensor_5}
\end{equation}
If we let $\kappa =\frac{-1}{\hbar^2} $, then $\hat{T}{\alpha\beta} = p^{\alpha}p_{\beta}$, i.e. energy momentum 
tensor is the products of two 4-dimensional momentum. Let $p_5 = m_0, p_4 = 0$, then 6-dimensional
engergy momentum tensor is: $\hat{T}{AB} = p^{A}p_{B}$.  One can also see that as we describe quantum field 
in {\it pure geometry}, $\frac{i}{\hbar}$ plays the similar {\it geometry} role as $\sqrt{8\pi G}$ 's 
role in 4-dimensional Einstein equation, where G is gravational constant. 

\section{Equations of 1-spin free particle} \label{OneSpin}

For 1-spin particle with mass $m_0 > 0 $, we let 
\begin{equation}
\hat{A}_{\alpha} = A_{\alpha}e^{im_{0} x_5}
\label{hatA}
\end{equation}
where where $m_{0}$ is rest mass of particle, $x_5$ is 6th dimension coordinate, 
and start from this section, we always choose $\hbar \equiv  1$.
Let 
\begin{equation}
\hat{A}_{5} = 0
\label{hatA_5}
\end{equation}
6-dimensional metric for 1-spin free particle become:
\begin{equation}
\left( \hat{g}_{AB} \right) = \left( \begin{array}{cc}
   g_{\alpha\beta} + \hat{A}_{\alpha} \hat{A}_{\beta} \; \; & \; \; 
       \hat{A}_{\alpha} \; \\
    \hat{A}_{\beta} \; \;                                & \; \; 
     1 \; \; \; \; \; \; \\
  \; \; \; \; \; \; \; \; \;  & \; \; \; \; \; \; \; \; \; \; \;  \; \;   -1  \end{array} \right) 
\label{6dMetric1S_m}
\end{equation}
Let $\hat{F}_{AB} \equiv \partial_{A} \hat{A}_{B} - 
\partial_{B} \hat{A}_{A}$, and A, B runs over 0,1,2,3,5. 
energy momentum tensor
\begin{equation}
\hat{T}_{AB} \equiv g_{AB} \hat{F}_{CD} \hat{F}^{CD}/4 - \hat{F}_{A}^{C} \hat{F}_{BC}
\label{1SpinT}
\end{equation}
where A,B run over (0,1,2,3,5).
so the $\alpha \beta-$, $\alpha 4- $, and 44-components of 
6-dimensional Einstein equations (\ref{5dEFE1}) become:
\begin{eqnarray}
G_{\alpha\beta} = \frac{1}{2} 
   \hat{T}_{\alpha\beta} \; , \; \; \; 
\nabla^{\alpha} \, \hat{F}_{\alpha\beta} - m_0^{2} \hat{A}_{\beta} = 0 \; \nonumber \\
\frac{1}{4}\hat{F}_{\alpha\beta} \hat{F}^{\alpha\beta} - \frac{1}{2} m_{0}^{2} \hat{A}_{\alpha} \hat{A}^{\alpha} = 0 \; \; \; 
\label{6dFieldEquns1S_m}
\end{eqnarray}
This is equations for 1-spin single particle with mass $>$ 0. 
As we see above, the particle of 1-spin obtains its mass from derivative of 6th dimension.

When $m_0 = 0$, metric (\ref{6dMetric1S_m}) becomes usual Kaluza metric at the case of $\phi = -1$. The second and 
third equation of equations (\ref{6dFieldEquns1S_m}) become
\begin{eqnarray}
\nabla^{\alpha} \, \hat{F}_{\alpha\beta} = 0 \;, \; \; \;
\frac{1}{4}\hat{F}_{\alpha\beta} \hat{F}^{\alpha\beta} = 0 \; 
\label{Maxwell_1}
\end{eqnarray}
The first equation of (\ref{Maxwell_1}) is Maxwell equation in vacuum and the second equation of (\ref{Maxwell_1}) is true 
for free photon (plane-wave).

\section{Equations of $\frac{1}{2}$-spin free particle} \label{halfSpin}

In section \ref{OneSpin} for 1-spin particle, we choose $\hat{A}_{5} = 0$, now we will see that if we 
let $\hat{A}_{5} <> 0$, we will get field equations for $\frac{1}{2}$ particle. We choose the 6-dimensional metric
for  $\frac{1}{2}$ particle as: 
\begin{equation}
\left( \hat{g}_{AB} \right) = \left( \begin{array}{cc}
   g_{\alpha\beta} + \hat{K}_{\alpha} \hat{K}_{\beta} \; \; & \; \; 
      \hat{K}_{\alpha} \; \; \; \; \; \;  \; \;  \hat{K}_{\alpha} \hat{K}_{5} \\
    \hat{K}_{\beta} \; \;  & \; \; 
     1 \; \; \; \; \; \;  \; \;   \hat{K}_5  \\ 
   \hat{K}_{5} \hat{K}_{\beta} \; \; & \; \;  \hat{K}_5 \; \; \; \;  \; \;  -1+\hat{K}_{5} \hat{K}_{5}  \end{array} \right) \;
\label{6dMetricHalfS}
\end{equation}
where $\hat{K}$ is 6-dimensional vector field in 6-dimensional time-space, with conditions
\begin{eqnarray}
\partial_5 \hat{K}_{A} = im \hat{K}_{A} \;, \; \; \hat{K}_5 <> 0 \;, \; \; \hat{K}_4 = 0 \;, \; \; \partial_4 \hat{K}_A = 0 
\end{eqnarray}
and 
\begin{equation}
\left( \hat{g}^{AB} \right) = \left( \begin{array}{cc}
   g^{\alpha\beta}  \; \;  & \; \; 
    - \hat{K}^{\alpha} \; \; \;  \; \; \; \\
   - \hat{K}^{\beta} \; \;  & \; \; 
     1+\hat{K}_{A}\hat{K}^{A} \; \; \; \;- \hat{K}^{5}\\
 \; \;  & \; \;  - \hat{K}^{5} \; \; \; \; \; \;   -1  \end{array} \right) 
\label{6dMetricHalfS_i}
\end{equation}

Let $\hat{E}_{AB} \equiv \partial_{A} \hat{K}_{B} - 
\partial_{B} \hat{K}_{A}$. 
Define energy momentum tensor for half spin particle: 
\begin{equation}
\hat{T}_{AB} \equiv g_{AB} \hat{E}_{CD} \hat{E}^{CD}/4 - \hat{E}_{A}^{C} \hat{E}_{BC}
\label{halfSpinT}
\end{equation}
so the AB-, A4-, and 44-components of 
6-dimensional Einstein equations (\ref{5dEFE1}) become:
\begin{eqnarray}
G_{AB} = \frac{1}{2} \hat{T}_{AB} \;, \; \; \; \;
\partial_{A} (\partial^{C} \hat{K}_{C}) - \partial^{C} \partial_{C} \hat{K}_{A} = 0   \nonumber \\
\frac{1}{4}\hat{E}_{AB} \hat{E}^{AB} = 0  
\label{halfspinEq}
\end{eqnarray}
To derive the equation above, we used the relation below:
\begin{equation}
\partial_{C} g^{CC} (\partial_{A} \hat{K}_{C} - \partial_{C} \hat{K}_{A} ) 
 = \partial_{A} (\partial^{C} \hat{K}_{C} ) - \partial^{C} \partial_{C} \hat{K}_{A} 
 \label{relation}
 \end{equation}

For free particle, it is reasonable to assume that each components of $\hat{K}$ satisfied plane-wave condition:
\begin{equation}
\partial^{C} \partial_{C} \hat{K}_{A} = 0
\label{planewave_1}
\end{equation}
The second equation of equations (\ref{halfspinEq}) become
\begin{equation}
\partial_{A} (\partial^{C} \hat{K}_{C})  =  0  \; \; \; \; for \, all \, A = 0,1,2,3,5
\label{halfSpinEq2}
\end{equation}
i.e. $\partial^{C} \hat{K}_{C} $ does not depended on $x_0,x_1,x_2,x_3,x_5$, 
 so it is reasonable to let
$\partial^{C} \hat{K}_{C} $ equals zero (Not likely equals a constant other than zero 
becaues $\hat{K}$ contains
a plane-wave function part of condition (\ref{planewave_1})). Together with the third equation of 
equations (\ref{halfspinEq}), we have
\begin{eqnarray}
\partial^{C} \hat{K}_{C}  = 0  \;, \; \; \; \; 
\hat{E}_{AB} \hat{E}^{AB} = 0   \; \; \; \; \;
\label{halfSpinEq3}
\end{eqnarray}
Now let 
\begin{eqnarray}
\hat{K_0} = Cg_{00}\phi_0 e^{im_0 x_5}   \; \; \; \;
\hat{K_1} = Cg_{11}\phi_3 e^{im_0 x_5}  \nonumber \\ 
\hat{K_2} = -iCg_{22} \phi_3 e^{im_0 x_5} \; \; \; \;
\hat{K_3} = Cg_{33}\phi_2 e^{im_0 x_5}  \nonumber \\ 
\hat{K_5} = Cg_{55}\phi_0 e^{im_0 x_5}   
\label{HalfSpinVector}
\end{eqnarray}
where C is constant to be determined, $g_{\alpha \alpha}$ is element of usual 4-dimensional metric (1,-1,-1.-1), $g_{55} = -1$. 
First equation of (\ref{halfSpinEq3}) becomes 
\begin{equation}
\partial_{0} \phi_0 + \partial_{1} \phi_3 - i \partial_{2} \phi_3 + \partial_3 \phi_2 + im_0 \phi_0 = 0
\label{ModifiedDirac}
\end{equation}
It is first Dirac equation in $x_3$ representation. The two equations in (\ref{halfSpinEq3}) plus normalization
condition of $\phi_0$, $\phi_2$, $\phi_3$ keep the solution unique, plus we need choose constant C to make the energy momentum
tensor $\hat{T}_{AB}$ reasonable. The solutions of  (\ref{halfSpinEq3}) are:
\begin{eqnarray}
\hat{K}_0 = C \sqrt{\frac{m_0+p_0 }{2m_0}} e^{-i p^A x_{A}}\;, \; \; \; \hat{K}_5 = - \hat{K}_0 \;, \;  \; \; \ \nonumber \\
\hat{K}_1 = -C\sqrt{\frac{m_0+p_0 }{2m_0}} \frac{p_1 + ip_2}{m_0+p_0} e^{-i p^A x_{A}}  \;, \; \; \; \;  \nonumber \\
\hat{K}_2 = iC\sqrt{\frac{m_0+p_0 }{2m_0}} \frac{p_1 + ip_2}{m_0+p_0} e^{-i p^A x_{A}}  \;, \; \; \; \;  \nonumber \\
\hat{K}_3 = -C\sqrt{\frac{m_0+p_0 }{2m_0}} \frac{p_3}{m_0+p_0} e^{-i p^A x_{A}} \; \;  \; \; \;  
\label{DComponents}
\end{eqnarray}
The corresponding $\phi_0$, $\phi_1 = 0$,
$\phi_2$, $\phi_3$ in relation (\ref{HalfSpinVector}) is the first solution of Dirac equation of $\frac{1}{2}$ spin particle with positive energy. 
We choose the constant C in equations (\ref{HalfSpinVector}) as
\begin{equation}
C = \frac{\sqrt{(m_0+p_0)2m_0}}{p_3 \sqrt{p_0}}
\label{Constant}
\end{equation}
Substitute (\ref{Constant}), (\ref{DComponents}) into (\ref{halfSpinT}), then:
\begin{equation}
\hat{T}_{ab} = -\frac{p_{a}}{p_0}p_{b}e^{\frac{-2i}{\hbar}p^{c}x_{c}} 
\label{TAB}
\end{equation}
The negative sign can be removed by adding a constant $\kappa$ in metric and let $\kappa = \frac{-1}{\hbar^2} $ as we did for
zero spin particle. In relativity, we know that $v_a = \frac{p_{a}}{p_0} $, so  
the result is just we expected: the energy momentum tensor becomes
the 5-dimensional momentum vector $(p_{\alpha},m_0)$ multiply speed and times the square of plane wave function.

If we let 
\begin{eqnarray}
\hat{K_0} = Cg_{00}\phi_1 e^{ im_0 x_5}  \;, \; \; \; \;
\hat{K_1} = Cg_{11}\phi_2 e^{ im_0 x_5}  \nonumber \\
\hat{K_2} = iCg_{22}\phi_2 e^{im_0 x_5}  \;, \; \; \; \;
\hat{K_3} = -Cg_{33}\phi_3 e^{im_0 x_5}  \nonumber \\
\hat{K_5} = Cg_{55}\phi_1 e^{im_0 x_5}  
\label{HalfSpinVector_2} 
\end{eqnarray}
We can get local time-space metric of half spin free particle with spin $-\frac{1}{2}$ and positive energy;
Easy to examine that, it is also satisfied both equations of (\ref{halfSpinEq3}) and Dirac equations.
Similarly we can derive the other two solutions for negative energy of Dirac equations.  As conclusion, equations
(\ref{halfspinEq}) is general field equations for particles with half integer spin. It is interesting seeing that:
equation (\ref{halfSpinEq3}) together with plane-wave condition (\ref{planewave_1}), we have:
\begin{equation}
\nabla^{A}(\hat{E}_{AB}) = 0   
\label{Maxwell_ex}
\end{equation}
where $ \hat{E} \equiv \partial_{A} \hat{K}_{B} - 
\partial_{B} \hat{K}_{A} $. It is similar to Maxwell equation. If we let $\hat{K}_5 = 0$, it becomes
the second equation of (\ref{6dFieldEquns1S_m}) for 1-spin particle; if we also let $m_0 = 0$, then the above equation
(\ref{Maxwell_ex}) turns into Maxwell equation. So 1-spin particle (both massless and mass $>0$), $\frac{1}{2}$-spin particle
are satisfied the same equations -- extended 6-dimensional Maxwell equation (\ref{Maxwell_ex}) in 6-dimensional time-space.
The field of $\frac{1}{2}$ particle becomes 6-dimensional vector field.

Compare the metrics of 1-spin particle with $\frac{1}{2}$-spin particle, 
the only difference is that one without 6th-componet 
of vector and the other with 6th-component. The metric of 1-spin particle has symmetry of 
4-dimensional time-space, the metric
of $\frac{1}{2}$-spin particle has symmetry of 5-dimensnional time-space (0,1,2,3,5).

\section{Equations of Electron-Photon Interaction} \label{Interaction}

The similarity between the metric of single electron and the metric of single photon 
makes us easier to combine eletron and photon into one metric, The metric $\hat{g}_{AB}$ of the coupling of
electron-photon is:
\begin{equation}
\left( \begin{array}{cc}
   g_{\alpha\beta} + \kappa^{2} \hat{B}_{\alpha} \hat{B}_{\beta} \; & \; 
    \kappa \hat{B}_{\alpha} \; \; \; \; \kappa^{2}\hat{B}_{\alpha} \hat{K}_{5}  \\
    \kappa \hat{B}_{\beta}  \; & \; 
     1 \; \; \; \; \;  \;  \; \;  \;  \kappa \hat{K}_5 \\
   \kappa^{2} \hat{K}_{5} \hat{B}_{\beta} \; & \; \; \; \kappa \hat{K}_5  \; \; \; \;  \;  -1+ \kappa^{2} \hat{K}_{5} \hat{K}_{5}  \end{array} \right) 
\label{6dMetricCoupl_e}
\end{equation}
where $\hat{B}_\alpha \equiv A_{\alpha}+\hat{K}_{\alpha} $;  $A_{\alpha}$ for 4-dimensional photon vector field, $\hat{K}$ for 
6-dimensional vector field (spinor in 4-dimension) which we defined in previous section for electron. 
$\kappa$ is constant for coupling with gravity; $\kappa = 4\sqrt{\pi G}$ where G is gravational constant.
To describe interactions between electron and photon, we impliment Klein's idea \cite{Kle26a} to make the 
derivative of 5th dimension $<>0$ , we can understand that in two ways: 1)
the interaction of Electron-Photon changed curvature of local time-space. 2) An Electron always associate with electromagnetic fields 
which cause non-zero derivative of 5th dimension.
Let
\begin{eqnarray}
\hat{K}_{A} \rightarrow \hat{K}_{A}e^{in\gamma x_4}   
\label{4thDeriv}
\end{eqnarray}
where $\gamma $ is a very small constant, n is integer, and there is no $\hat{K_4}$. Then $\partial_4 \hat{K}_{A} = i \gamma \hat{K}_{A} $.
Because the derivative of 5th dimension $<> 0$, 5th dimensional energy momentum tensor is not zero.

Using (\ref{6dMetricCoupl_e}), (\ref{4thDeriv}), through considerable algebra, the AB-, and 44-components of 
6-dimensional Einstein equations (\ref{5dEFE1}) become:
\begin{eqnarray}
G_{AB} = \frac{\kappa^2}{2} \hat{T}_{AB} \;, \; \; \; \;
\frac{1}{4}\hat{L}_{AB} \hat{L}^{AB} = \hat{T}_{44} 
\label{halfspinEq_coupl}
\end{eqnarray}
and the A4-components of Einstein equations can be write into 3 separate pieces:
\begin{eqnarray}
D^{C} \hat{K}_{C} = 0 \;, \; \; \; D^{C} D_{C} \hat{K}_{A} = 0 \;, \;\nonumber \\
\nabla^{\alpha} \, \hat{F}_{\alpha A} = 0 
\label{halfspinEq_coup2}
\end{eqnarray}
where $D_{\alpha} =\partial_{\alpha} - in\gamma A_{\alpha} $.
We extended photon field
to 6-dimension with $A_4 = 0, A_5 = 0$; $F_{AB} \equiv \partial_{A} A_{B} - \partial_{B} A_{A}$, 
$\hat{L}_{AB} \equiv \partial_{A} A_{B} - 
\partial_{B} A_{A} + D_{A}\hat{K}_{B} - D_{B} \hat{K}_{A}$, energy momentum tensor of electron coupling with photon becomes:
\begin{eqnarray}
\hat{T}_{AB} \equiv \frac{g_{AB}}{4}\hat{L}_{CD}\hat{L}^{CD} - \hat{L}_{A}^{C} \hat{L}_{BC} + \nonumber\\
     + in\gamma(\hat{B}_{B}\hat{K}^C F_{AC} + \hat{B}_{A}\hat{K}^C F_{BC}) \nonumber \\
     + in\gamma (D_{A}\hat{K}_{B} + D_{B}\hat{K}_{A}) 
\label{halfSpinT_e}
\end{eqnarray}
where we dropped $\gamma^2$ items in energy momentum tensor since $\gamma$ is small; $\hat{E}_{AB} \equiv \partial_{A} \hat{K}_{B} - 
\partial_{B} \hat{K}_{A}$. 

If let $n=1, \gamma = -e$, e is charge of electron, and use relation from (\ref{HalfSpinVector}), 
then first equation of (\ref{halfspinEq_coup2}) becomes:
\begin{eqnarray}
(\partial_{0} + ie A_0) \phi_0 + (\partial_{1} +ie A_1) \phi_3 - i (\partial_{2}+ie A_2) \phi_3 \; \; \; \; \nonumber \\  
 + (\partial_3 + ie A_3) \phi_2 + im_0 \phi_0 = 0
\label{ModifiedDirac_2}
\end{eqnarray}
This is Dirac equation for electron-photon interaction. 
The second equation of (\ref{halfspinEq_coup2}) is Klein-Gordon equation in electromagnetic field that
each components of electron field should satisfied individually. The 3rd equation of (\ref{halfspinEq_coup2})
is Maxwell equation.

Now let's look at energy momentum tensor (\ref{halfSpinT_e}). The first two parts are combined energy momentum tensor of electron
and photon. The third item $ie\hat{B}_{B}\hat{K}^C F_{AC}$ is related to Lorenz force. The third item is electric current density.

6-dimensional action  is
\begin{equation}
S = -\int \sqrt{-\hat{g}} \hat{R}  
     \; d^6 x \; \; \; 
\label{6dAction}
\end{equation}
where $\hat{R} = \hat{g}^{AB} \hat{R}_{AB} $ (A,B runs over 0,1,2,3,4,5)
is 6-dimensional Ricci scalar. The metric determinant $\hat{g}$ reduces in the simple manner:
\begin{equation}
\hat{G} = det(\hat{g}_{AB}) = - det(g_{\alpha\beta})
\label{metric_det}
\end{equation}
Using (\ref{6dMetricCoupl_e}), (\ref{halfspinEq_coupl}) and (\ref{halfspinEq_coup2}),
we also assume $\int dx_5 dx_6 = 1 $, then
\begin{equation}
S = -\int \sqrt{-\hat{g}}( R + \frac{1}{4} \hat{L}_{AB} \hat{L}^{AB} )
     \; d^4 x \; \; \; 
\label{6dAction_coupl}
\end{equation}
where R is usual 4-dimensional Ricci scalar. The above equation contains a gravity part and kinetic energy part.
for interaction between electron and photon. 

\section{Discussions and Conclusions} \label{SUM}

The kinetic energy part of (\ref{6dAction_coupl}) $\frac{1}{4} \hat{L}_{AB} \hat{L}^{AB}$ 
contains the combination of electron field and 
photon field. Actually, metric (\ref{6dMetricCoupl_e}) can be used to describe: 

1) {\it Eletron associate with its own static electromagnetic field}. Eletron is always accompany with photon (static
electromagnetic fields), that's why it can absort photon to gain energy, emits phton and lost energy, annihilation with
positron and turns into two photons, so it is reasonable to say that there is 
always $A_{\alpha}$ field in electron's local geometry metric.

2) {\it Electron absorb a photon and gains energy}. If a free photon $A_0$ at time $t_0$, location $X(x_{01},x_{02},x_{03})$ which
local geometry metric is (\ref{6dMetric1S_m}), it is absorted by an electron at time t, then the local metric at time t becomes
metric (\ref{6dMetricCoupl_e}) with $\hat{B^{\prime}} = A^{\prime}_0 + A_{e} + \hat{K}$, where photon $A_0$ turns to $A^{\prime}_0$ after 
interaction and $A_{e}$ is initial photon contained in electron, and finally electron gain the energy.

3) {\it The intermediate state of a photon interacts with electron}. If a free photon $A_0$ at time $t_0$, location $X(x_{01},x_{02},x_{03})$ which
local geometry metric is (\ref{6dMetric1S_m}), it interacts with electron at time t, then the local metric at time t becomes
metric (\ref{6dMetricCoupl_e}) with $\hat{B^{\prime}} = A^{\prime}_0 + A_{e} + \hat{K}$, after interaction, the photon and electron
will be separate. Of couse the energy and momentum of both particle changed.

In the discussion above, we only dealing with curvature changes. We do not need use any ``virtual'' photons.

One can see from section \ref{OneSpin}, U(1) field obtained its mass from derivative of 6th dimension without losing symmetry, so 
we could avoid Higgs mechanism. If we treate interactions as changes of geometry curvature of local time-space, we could 
avoid so called ``vacuum effects'' in current QED theory, the ``vaccum effects'' could be interpreted as local curvature changes too.

In addition, here we see a symmetry between gravity and electromagnetism: Mass is the source of gravity, it comes from the derivative 
of 6th dimension; Charge is the source of electromagnetic interaction, it comes from the derivative of 5th dimension.
We can also use Klein's original idea that, the 5th and 6th have circular topology and periodic conditions, that can make
quantization of electric charge and quantization of mass (or energy).

As the conclusion, in this paper, we derived above equations directly from Einstein equations under 6-dimensional time-space metric. 
The energy momentum tensor is
the products of extended 6-dimensional momentum vector $p_A$ (where $p_5 = m_0$, $p_4 = 0$, no 5 dimensional momentum);
Planck constant $\hbar$ plays the similar role as the role of gravational constant G in gravity equations. 
The interactions between photon and electron can be treated as the effects of local time-space geometry curvature changing.
It indicates the pure geometry nature of quantum fields.
Indeed, by using metric (\ref{6dMetric_0}), one can find that the geodesic path of single free
 particle in 6-dimensional time-space is exact plane-wave function, the interpretion of basic quantum physics by using
 6-dimensional time-space is discussed in another paper \cite{xchen}.

\end{document}